\documentstyle[12pt,fleqn]{article}
\addtocounter{section}{-1}
\begin{document}

\begin{center}

\vspace{2cm}

{\large {\bf {
Possible charge ordered states\\
in BN and BCN nanotubes and nanoribbons
} } }

\vspace{1cm}

{\rm Kikuo Harigaya\footnote[1]{E-mail address: 
\verb+k.harigaya@aist.go.jp+; URL: 
\verb+http://staff.aist.go.jp/k.harigaya/+}}

\vspace{1cm}

{\sl Nanotechnology Research Institute, AIST, 
Tsukuba 305-8568, Japan}\footnote[2]{Corresponding address}\\
{\sl Synthetic Nano-Function Materials Project, 
AIST, Tsukuba 305-8568, Japan}

\vspace{1cm}

(Received~~~~~~~~~~~~~~~~~~~~~~~~~~~~~~~~~~~)
\end{center}

\vspace{1cm}

\noindent
{\bf Abstract}\\
Electronic states in boron-nitride and boron-carbon-nitride
nanoribbons with zigzag edges are studied using the extended 
Hubbard model with nearest neighbor Coulomb interactions.   
The charge and spin polarized states are considered, and
the phase diagram between two states is obtained.
Next, the electric capacitance is calculated in order to 
examine the nano-functionalities of the system.  Due to the 
presence of the strong site energies, the charge polarized 
state overcomes the spin polarized states, giving the large 
difference of the phase diagram in comparison with that of 
graphite system.   The electronic structures are always like 
of semiconductors.  The capacitance calculated for the 
charge polarized state with the realistic values of the Coulomb 
interactions is inversely proportional to the ribbon width, 
owing to the presence of the charge excitation energy gap.

\vspace{1cm}

\pagebreak

\section{Introduction}

Nano-carbon (C) materials and hetero-materials including borons 
(B) and nitrogens (N) have been attracting much attention 
both in the fundamental science and in the interests of
application to nanotechnology devices.$^{1,2)}$  Their physical 
and chemical natures change variously depending on geometries.$^{1-3)}$.  
In carbon nanotubes, diameters and chiral arrangements 
of hexagonal pattern on tubules decide whether they are 
metallic or not.$^{1,2)}$

In nanographites, the edge atoms strongly affect the electronic 
states,$^{3)}$ and there are nonbonding molecular orbitals localized 
mainly along the zigzag edges.  Recently, we have studied the 
competition between the spin and charge orderings due to the 
on-site and nearest neighbor Coulomb interactions.$^{4)}$  The 
nearest neighbor Coulomb interaction stabilizes the charge 
polarized (CP) state with a finite electric dipole moment in 
zigzag ribbons, and it competes with the spin polarized (SP) 
state.  It has been discussed that the transverse electric 
field might induce the first order phase transition from the 
SP state to the CP state.

In view of these findings, it is interesting to perform
further theoretical investigations for BN (BCN) nanotubes 
and nanoribbons.  The BN systems are hetero-atom materials,
where B and N are located alternatively on the honeycomb 
lattice.  The BN plane and nanotubes are intrinsic insulators
with the energy gap about 4 eV.$^{5,6)}$  Low energy excitation
properties have not been investigated so much.  We will discuss
interplay between CP and SP states, even though there remain
limitations of theories of a single orbital only.  We note
that photogalvanic effects have been discussed using the
single orbital theory.$^{7)}$  Our idea is similar to this study,
and we will introduce site energies at the B and N sites.
The other system, BCN nanotubes and nanoribbons, is a model
material, where B atoms are arrayed along one edge and N
atoms along the other edge. The experiments of the low 
concentration limit of B and/or N doping into carbon nanotubes 
have sometimes suggested accumulation of impurity atoms 
at edge sites.$^{8,9)}$  The formation of zigzag nanotubes is 
favored.  We will consider electronic properties of zigzag
BCN nanoribbons, too.

We will discuss that spin excitations tend to be suppressed
due to the presence of the large energy gap.  This gives 
rise to the enlarged region of the CP state in the phase
diagrams of the BN and BCN systems.  This property is 
largely in contrast to the dominant SP state of the
zigzag graphite ribbons.$^{4)}$

The next interest of this paper is how such the electronic properties
appear in measurable quantities.  We will consider electric
capacitance because the CP state is accompanied with dielectric
moment with respect to the charge orders.  We model the
periodic zigzag ribbon as a ``nano-size condenser", where
a set of the B edge atoms are regarded as a positive
electrode, a set of the N edge atoms as a negative
electrode, and a material as a spacer is present between
the positive and negative electrodes.  The ``differential
capacitance" will be calculated as a response of charge
with respect to the weak static field applied to the system
in the transverse direction.  The similar
idea of the calculations has been used in the study using
the first principle methods.$^{10)}$  We find that the calculated 
capacitance is inversely proportional to the distance 
between the positive and negative electrode.  This behavior 
could be understood by the presence of an energy gap for 
charge excitations.

This paper is organized as follows.  In section 2, we explain
the extended Hubbard model, and give the calculation method.  
In section 3, we give results for the BN systems.  In section 4, 
calculations of the BCN systems are reported.  The paper is 
closed with a short summary in section 5.

\section{Model and method}

Figure 1 illustrates the geometry of zigzag nanoribbons,
where $N$ and $L$ are the 
width and length of the ribbon, respectively.  Figure
1 (a) shows the zigzag BN ribbons, and Fig. 1 (b) shows
the zigzag BCN ribbons.  The periodic boundary condition 
is set along the $y$-axis parallel to the zigzag-lines.  
Since the zigzag ribbon is a bipartite lattice, $A$ (boron) and 
$B$ (nitrogen) sites are assigned by the filled and open 
circles in Fig. 1 (a), respectively.  All the twofold 
coordinated sites in the lower and upper zigzag edges 
belong to the $A$ and $B$ sublattices, respectively, 
at which edge states are mainly localized.
In Fig. 1 (b), the B atoms are present along the upper
zigzag edge sites, while there are the N atoms along
the lower edges.  The inner part is composed of C 
atoms.  The closed, shaded, and open circles are for
B, C, and N atoms, respectively.  The experiments of 
the low concentration limit of B and/or N doping into 
carbon nanotubes have sometimes suggested accumulation 
of impurity atoms at edge sites.$^{8)}$  The formation of 
zigzag nanotubes is favored.  Therefore, we choose 
this structure as a model system.

We treat a half-filled $\pi$-electron system on the 
zigzag ribbon using the extended Hubbard Hamiltonian 
with the on-site $U$ and nearest-neighbor $V$ Coulomb 
interactions.  The model is as follows:
\begin{eqnarray}
H &=& E_{\rm B} \sum_{i \in {\rm B},\sigma} c_{i,\sigma}^\dagger c_{i,\sigma}
+ E_{\rm N} \sum_{i \in {\rm N},\sigma} c_{i,\sigma}^\dagger c_{i,\sigma}
\nonumber \\
&-& t \sum_{\langle i,j \rangle,\sigma}
( c_{i,\sigma}^\dagger c_{j,\sigma} + {\rm h.c.} )
+ U \sum_{i} 
(c_{i,\uparrow}^\dagger c_{i,\uparrow} - \frac{n_{\rm el}}{2})
(c_{i,\downarrow}^\dagger c_{i,\downarrow} 
- \frac{n_{\rm el}}{2}) \nonumber \\
&+& V \sum_{\langle i,j \rangle}
(\sum_\sigma c_{i,\sigma}^\dagger c_{i,\sigma} - n_{\rm el})
(\sum_\tau c_{j,\tau}^\dagger c_{j,\tau} - n_{\rm el}),
\end{eqnarray}
where $E_{\rm B}$ and $E_{\rm N}$ are the site energies at the B and N 
sites, respectively; the sum with $i \in {\rm B}$ and $i \in {\rm N}$ 
are taken over the B and N atoms, respectively; $c_{i,\sigma}$ 
annihilates a $\pi$-electron of spin $\sigma$ at the $i$th site; 
$t$ ($> 0$) is the hopping integral between the nearest neighbor 
$i$th and $j$th sites; the sum with $\langle i,j \rangle$ is taken for 
all the pairs of the nearest neighbor sites;  $n_{\rm el}$ is 
the average electron density of the system.  We adopt the 
standard Unrestricted Hartree-Fock approximation 
to this model:$^{4,11)}$
\begin{equation}
c_{i,\uparrow}^{\dagger}c_{i,\uparrow} 
 c_{i,\downarrow}^{\dagger}c_{i,\downarrow} 
\Rightarrow \langle c_{i,\uparrow}^{\dagger}c_{i,\uparrow}\rangle
c_{i,\downarrow}^{\dagger}c_{i,\downarrow}
+c_{i,\uparrow}^{\dagger}c_{i,\uparrow}
\langle c_{i,\downarrow}^{\dagger}c_{i,\downarrow}\rangle
-\langle c_{i,\uparrow}^{\dagger}c_{i,\uparrow}\rangle
\langle c_{i,\downarrow}^{\dagger}c_{i,\downarrow}\rangle.
\end{equation}

In order to examine nano-functionalities as ``nano-size 
condensers", electric capacitance of the zigzag ribbons 
is calculated.   We assume that one set of B atoms along the 
upper zigzag edge is regarded as a positive electrodes.  
The other set of N atoms along the lower zigzag edge is 
regarded as a negative electrodes.  The absolute value 
of the net variation of the accumulated charge is divided
by the strength of the small applied voltage, and the 
capacitance is obtained.  The quantity $Q_0$ is the net 
charge over the $L/2$ B or N atoms at the zigzag edge sites, 
when the static electric field is absent.  The part $dQ$ 
is the change of the net charge with respect to the small 
field which is parallel to the $x$-axis of Fig. 1 (a).  
The capacitance $C$ is calculated using the relation of 
polarizability $dQ = C dV$, where $dV$ is the change of 
voltage due to the static electric field.  The similar
method has been used for the calculation of the capacitance
using the first principle techniques recently.$^{10)}$  Here, we 
assume that the bond length between carbons is 1.45~\AA.

\section{Zigzag BN nanotubes and nanoribbons}

We will discuss electronic properties of the zigzag
BN nanotubes and nanoribbons.  The lattice structure
is shown in Fig. 1 (a).  The site energies 
at the B and N are taken to be $E_{\rm B} = +t$ and $E_{\rm N} 
= -t$.  The realistic magnitude $t \sim$ 2 eV gives the energy 
gap $2t \sim$ 4 eV, which has been reported in the band calculations.$^{8,9)}$
The similar choice of site energies in the single-orbital
tight binding model has been done for the random doping 
of BCN alloys in the literature.$^{12)}$  The total electron 
number is same with that of the site number.  This ensures
the charge neutrality of the system.

Figure 2 (a) shows the typical example of charge density 
distribution of the CP state for $U=1t$ and $V=0$, and 
Fig. 2 (b) displays the $z$-component of spin density 
distribution of the SP state for $U=4t$ and $V=0$.
The size of the zigzag ribbon is with $4 \times 20$ sites.  
The filled and open circles show positive and negative densities.  
Their radii are proportional to the magnitudes of the 
charge or spin densities: the maximum is 0.60 in Fig. 2 (a) 
and 0.35 in Fig. 2 (b).  Due to the strong site energies,
there are alternations of electron number density which
are extend over the system in Fig. 2 (a).  Furthermore,
at the B sites along the upper edge, the positive charge
density is enhanced than that of the inner part of the
ribbon.  At the N sites along the lower edges, the negative
density becomes enhanced similarly.  As the interaction
$U$ becomes stronger, the system changes into the SP state via phase 
transition. Fig. 2 (b) shows the spin density distributions.  The
alternations of spins are quite strong owing to the
very large interaction parameter.  The absolute values
of the spin density are enhanced along the two edges.
Such the modulations of the CP and SP near the edges are 
the result of the wavefunctions localized along the edge 
sites, as we have discussed for graphite ribbons.$^{4)}$

The total energies of the CP and SP solutions are compared,
and the phase diagram of the stable state is given in Fig. 3.
The calculation has been done for $N=4$ and $L=40$.
The phase boundary, denoted by the bold line, indicates
the first order phase transition.  The dashed line is the
strong correlation limit, $V = N(U + 2E_{\rm N})/(3N-1) = 4(U-2t)/11$,
which can be obtained by taking the energies of the two states
as equal.  The bold line approaches to the dashed line,
when $U$ and $V$ become larger.  The strong site energy 
difference gives rise to huge charge polarizations, as shown
in Fig. 2 (a).  The region of the CP state extends 
in the phase diagram, and the SP state is highly suppressed.  
The realistic values of the interactions might be near to
the regions, $0 < U < 3t$ and $V \sim 0$, which correspond 
to CP states in the phase diagram.  For the nanographite
ribbon,$^{4)}$ the phase boundary crosses the origin 
$(U,V) = (0,0)$.  So, the SP state has been found in these
regions.  The strong site energies of the BN system
have resulted in the quite remarkable change of ordered states 
between the graphite and BN systems.

Figure 4 shows the electric capacitance of the zigzag BN
ribbons of $L=20$ by changing the ribbon width.  As the
ribbon becomes wider, the system develops into the zigzag
(10,0) nanotube.  The raw value of the capacitance (a)
and the inverse (b) are plotted.  We take three parameter
sets of Coulomb interactions.  The system is in the CP 
state for these parameters.  All the curves exhibit the 
almost inversely proportional behaviors of the capacitance 
as functions of the ribbon width.  There is a huge electronic 
gap due to the strong site energy difference.  The system 
is a semiconductor intrinsically.  The strong charge excitation 
energy gap results in the inversely proportional behaviors.
Even though the present capacitance is the ``differential
capacitance", the dependence on the spacer width is similar
to that of the classical parallel electrode condenser.
Such the analogy seems interesting in view of the fact
that the spacer of the ``nano-condenser" is made of the
intrinsic insulator of the BN system.

\section{Zigzag BCN nanotubes and nanoribbons}

In this section, we report electronic properties of the zigzag
BCN nanotubes and nanoribbons, comparing with those of
the BN systems.  The lattice structure of the BCN systems
is shown in Fig. 1 (b).  In the present calculations, the 
site energies at the B and N are taken to be $E_{\rm B} = +0.8t$ 
and $E_{\rm N} = -0.8t$.  Because the most part of the system is
composed of carbons, we take $t \sim 2.5$ eV.  So, the site
energies at B and N take the same value as used in the previous
section.  The total electron number is same with that of 
the site number again.   Note that we have checked that
the calculated results change very weakly even when we take 
the site energies $E_{\rm B} = +t$ and $E_{\rm N} = -t$,
indicating that the slight change of the site energies
is not important for interpretations of the calculated results.

Figure 5 (a) shows the typical example of charge density 
distribution of the CP state for $U=1t$ and $V=0$, and 
Fig. 5 (b) displays the $z$-component of spin density 
distribution of the SP state for $U=4t$ and $V=0$.
The maximum of the charge density is 0.41 in Fig. 5 (a), 
and that of the spin density is 0.38 in Fig. 5 (b).  
The polarization of charge is localized along the upper
and lower edge sites in Fig. 5 (a).  It is extended over
the system in Fig. 2 (a).  The change of the distribution
patterns comes from the different kinds of atoms in the
inner region of the nanoribbon.  Fig. 5 (b) shows the 
spin density distributions of the SP state.  The
alternations of spins are quite strong, as we have
found in Fig. 2 (b).  The spin polarization is less 
dependent on the site energy difference, in contrast
to the charge polarization distribution.  Similarly,
spatial enhancement of the CP and SP along the two
edges is obtained in Figs. 5 (a) and (b), too.
This is the effect due to the presence of the edge states.

Figure 6 shows the phase diagram between the CP
and SP states, for the system with $N=4$ and $L=40$.
The phase boundary means the first order phase transition.
The dashed line, the strong correlation limit
$V = (N U + E_{\rm N})/(3N-1) = (4U-0.8t)/11$, approaches to 
the bold line, when $U$ and $V$ become larger.  
We find the enhanced area of the CP state, as in Fig. 3
of the BN ribbon.  The system will exhibit charge polarizations
for the realistic values of the interactions, too.
The decrease of the slope of the dashed line between
Figs. 3 and 6 is due to the decrease of the number
of B and N atoms, of course.

Finally, we look at the electric capacitance of the
BCN system.  The calculations are done for the system
size $L=20$, and the results are shown in Fig. 7.
We take two parameter sets of Coulomb interactions.  
The system is in the CP state for these parameters.
We obtain the almost inversely proportional behaviors 
of the capacitance as functions of the ribbon width again.
This is due to the strong charge excitation energy gap.
Quantitatively, the magnitude is larger than that
of Fig. 4, due to the fact that the inner region of
the ribbon is composed of carbon, and the local
energy gap is smaller than that of the BN system case.

\section{Summary}

The charge and spin polarized states have been discussed
using the extended Hubbard model for BN and BCN nanotubes 
and ribbons.  Next, the electric capacitance has been 
calculated in order to test the nano-functionalities of
the system.  Due to the presence of the strong site
energies, the CP state overcomes the SP states,
giving the large difference of the phase diagram
in comparison with the calculations of graphite system.  
The electronic structures of the BN and BCN systems are 
always like of semiconductors.  The capacitance calculated 
for the CP state with the realistic values of the Coulomb 
interactions is inversely proportional to the distance 
between the positive and negative electrodes, reflecting 
the presence of the charge excitation energy gap.

\begin{flushleft}
{\bf Acknowledgments}
\end{flushleft}

\noindent
This work has been supported by NEDO 
under the Nanotechnology Program.

\pagebreak
\begin{flushleft}
{\bf References}
\end{flushleft}

\noindent
$1)$ S. Iijima, Nature {\bf 354}, 56 (1991).\\
$2)$ R. Saito, G. Dresselhaus, and M. S. Dresselhaus, 
``Physical Properties of Carbon Nanotubes'', 
(Imperial College Press, London, 1998).\\
$3)$ M. Fujita, K. Wakabayashi, K. Nakada, 
and K. Kusakabe, J. Phys. Soc. Jpn. {\bf 65}, 1920 (1996).\\
$4)$ A. Yamashiro, Y. Shimoi, K. Harigaya, and K. Wakabayashi,
Phys. Rev. B {\bf 68}, 193410 (2003).\\
$5)$ A. Rubio, J. L. Corkill, and M. L. Cohen,
Phys. Rev. B {\bf 49}, 5081 (1994).\\
$6)$ X. Blase, A. Rubio, S. G. Louie, and M. L. Cohen,
Europhys. Lett. {\bf 28}, 335 (1994).\\
$7)$ P. Kr\'{a}l, E. J. Mele, and D. Tom\'{a}nek,
Phys. Rev. Lett. {\bf 85}, 1512 (2000).\\
$8)$ J. C. Charlier {\sl et al.}, Nano Lett. {\bf 2},
1191 (2002).\\
$9)$ M. Terrones {\sl et al.}, Mater. Today {\bf 7} (10), 30 (2004).\\
$10)$ N. Nakaoka and K. Watanabe, Eur. Phys. J. D {\bf 24},
397 (2003).\\
$11)$ K. Wakabayashi and K. Harigaya, 
J. Phys. Soc. Jpn. {\bf 72}, 998 (2003).\\
$12)$ T. Yoshioka, H. Suzuura, and T. Ando,
J. Phys. Soc. Jpn. {\bf 72}, 2656 (2003).\\

\pagebreak

\begin{flushleft}
{\bf Figure Captions}
\end{flushleft}

\mbox{}

\noindent
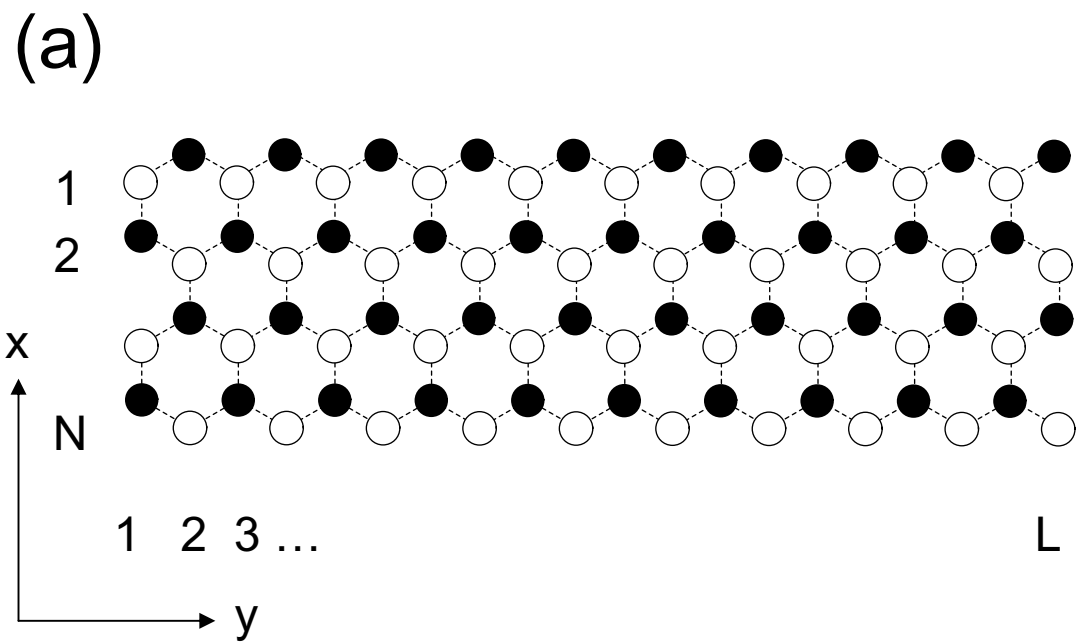
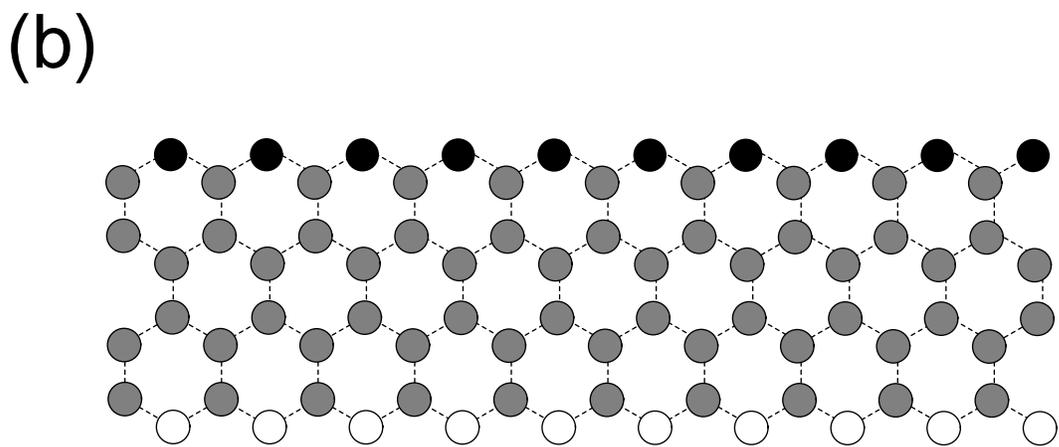
Fig. 1. Schematic structures of the (a) BN and (b) BCN ribbons 
with zigzag edges.  The filled, shaded, and open circles are 
B, C, and N sites, respectively.  The number of zigzag lines,
which are parallel to the $y$-axis, is denoted as $N$.  The
total number of atoms along the zigzag line is $L$.  Periodic 
boundary condition is imposed to the $y$-direction.  The number
of the B atoms along the upper edge and that of the N atoms
along the lower edge are $L/2$.

\mbox{}

\noindent
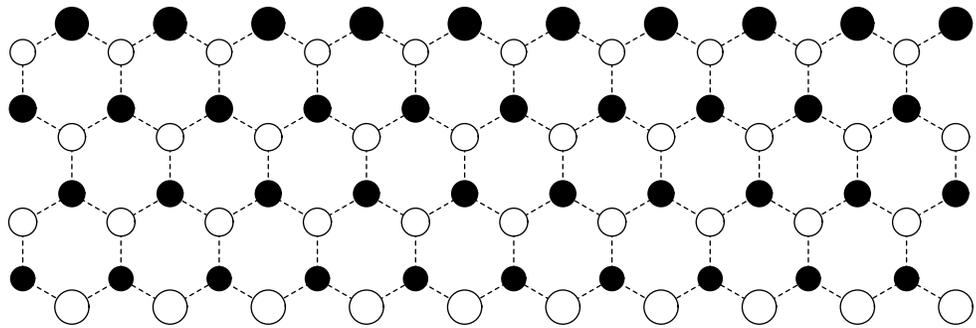
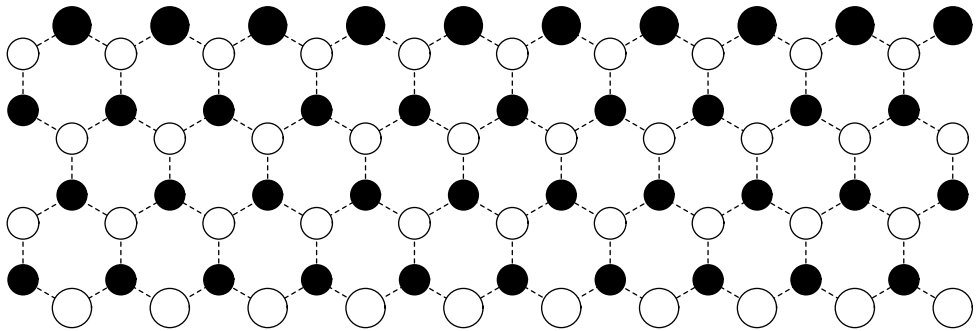
Fig. 2.  (a) Charge density distribution of the 
charge-polarized (CP) state for $U=1t$ and $V=0$, 
and (b) the $z$-component of spin density distribution 
of the spin-polarized (SP) state for $U=4t$ and $V=0$, 
on the zigzag BN ribbon with $4 \times 20$ sites.  The 
filled and open circles show positive and negative densities.  
Their radii are proportional to the magnitudes of the 
charge or spin densities: the maximum is 0.60 in (a) 
and 0.35 in (b).

\mbox{}

\noindent
Fig. 3. The phase diagram on the $U$-$V$ plane of the zigzag 
BN ribbon with $4 \times 40$ sites.  The bold line is the
phase boundary, and the dashed line is the strong correlation
limit.

\mbox{}

\noindent
Fig. 4.  The electric capacitance calculated 
for the CP state of the zigzag BN ribbons at $U = 0$,
$1t$, and $2t$ with $V=0$.  The ribbon length is $L=20$.  
The magnitudes of (a) the capacitance and (b) its inverse
are plotted against the ribbon width in the scale of~\AA.

\mbox{}

\noindent
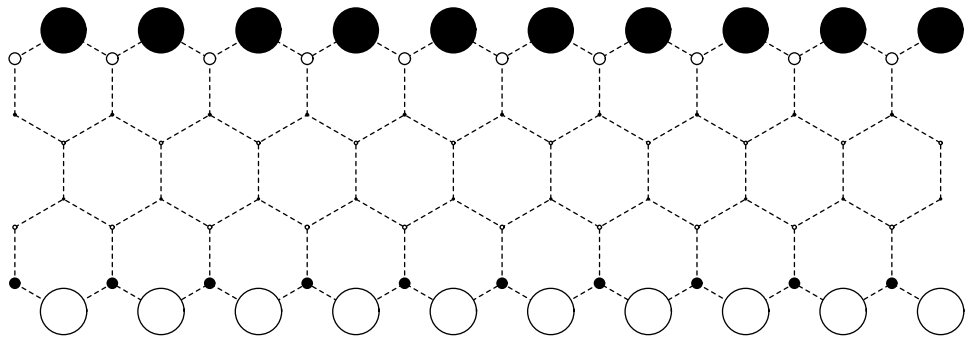
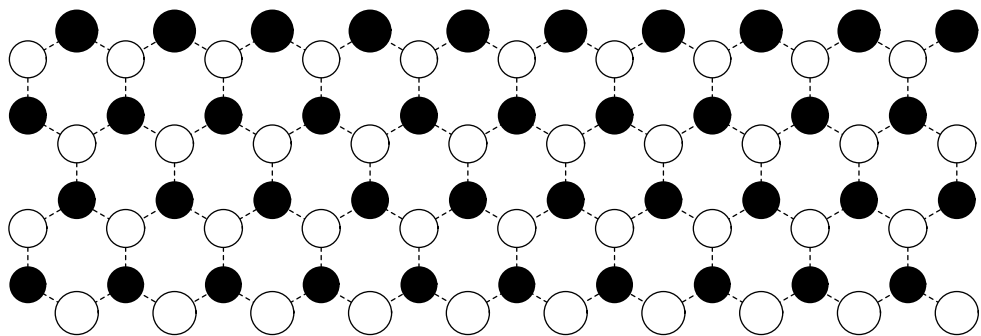
Fig. 5.  (a) Charge density distribution of the 
charge-polarized (CP) state for $U=1t$ and $V=0$, 
and (b) the $z$-component of spin density distribution 
of the spin-polarized (SP) state for $U=4t$ and $V=0$, 
on the zigzag BCN ribbon with $4 \times 20$ sites.  The 
filled and open circles show positive and negative densities.  
Their radii are proportional to the magnitudes of the 
charge or spin densities: the maximum is 0.41 in (a) 
and 0.38 in (b).

\mbox{}

\noindent
Fig. 6. The phase diagram on the $U$-$V$ plane of the zigzag 
BCN ribbon with $4 \times 40$ sites.  The bold line is the
phase boundary, and the dashed line is the strong correlation
limit.

\mbox{}

\noindent
Fig. 7. The electric capacitance calculated 
for the CP state of the zigzag BCN ribbons at $U = 0$
and $1t$ with $V=0$.  The ribbon length is $L=20$.  
The magnitudes of (a) the capacitance and (b) its inverse
are plotted against the ribbon width in the scale of~\AA.

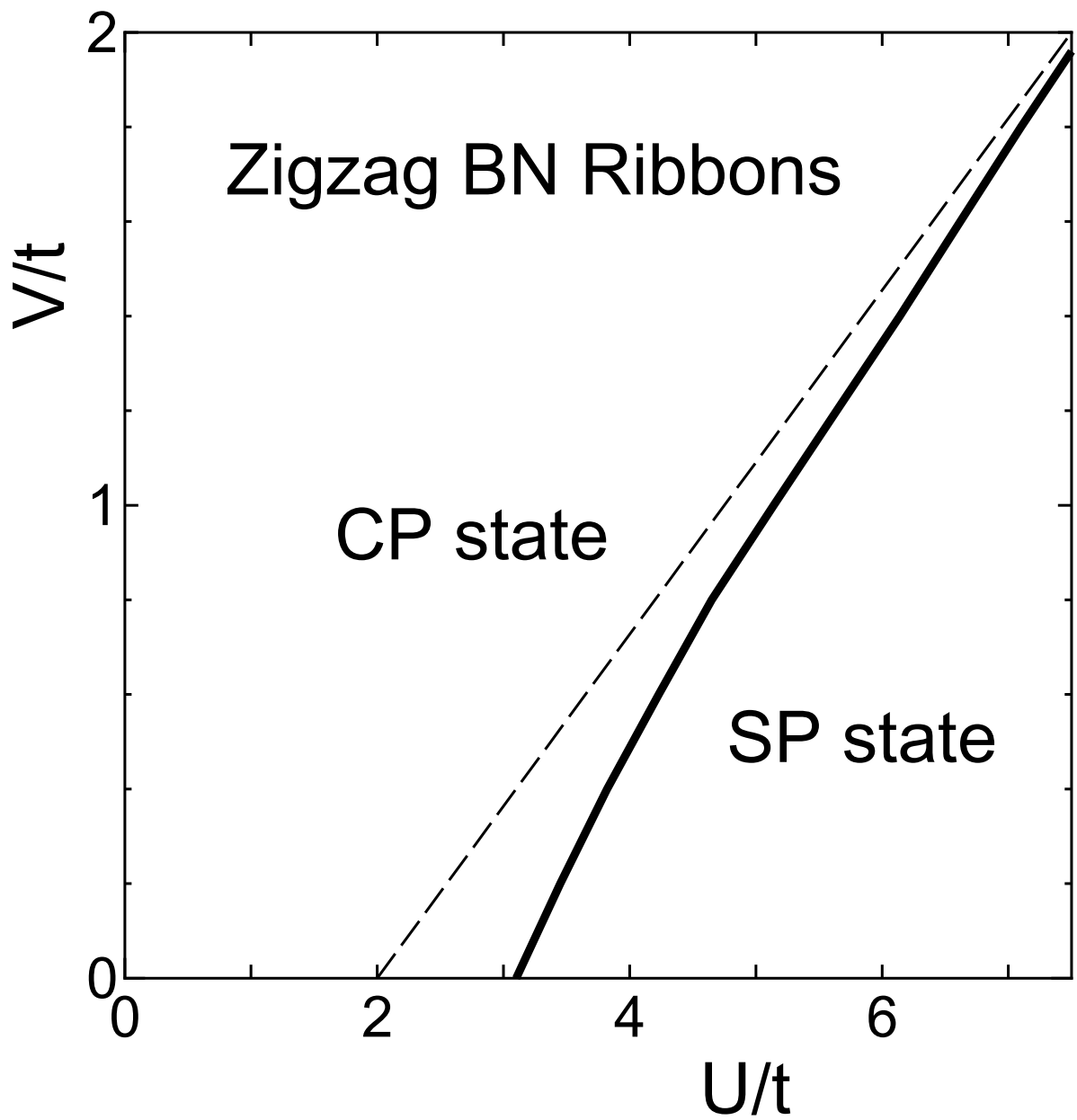
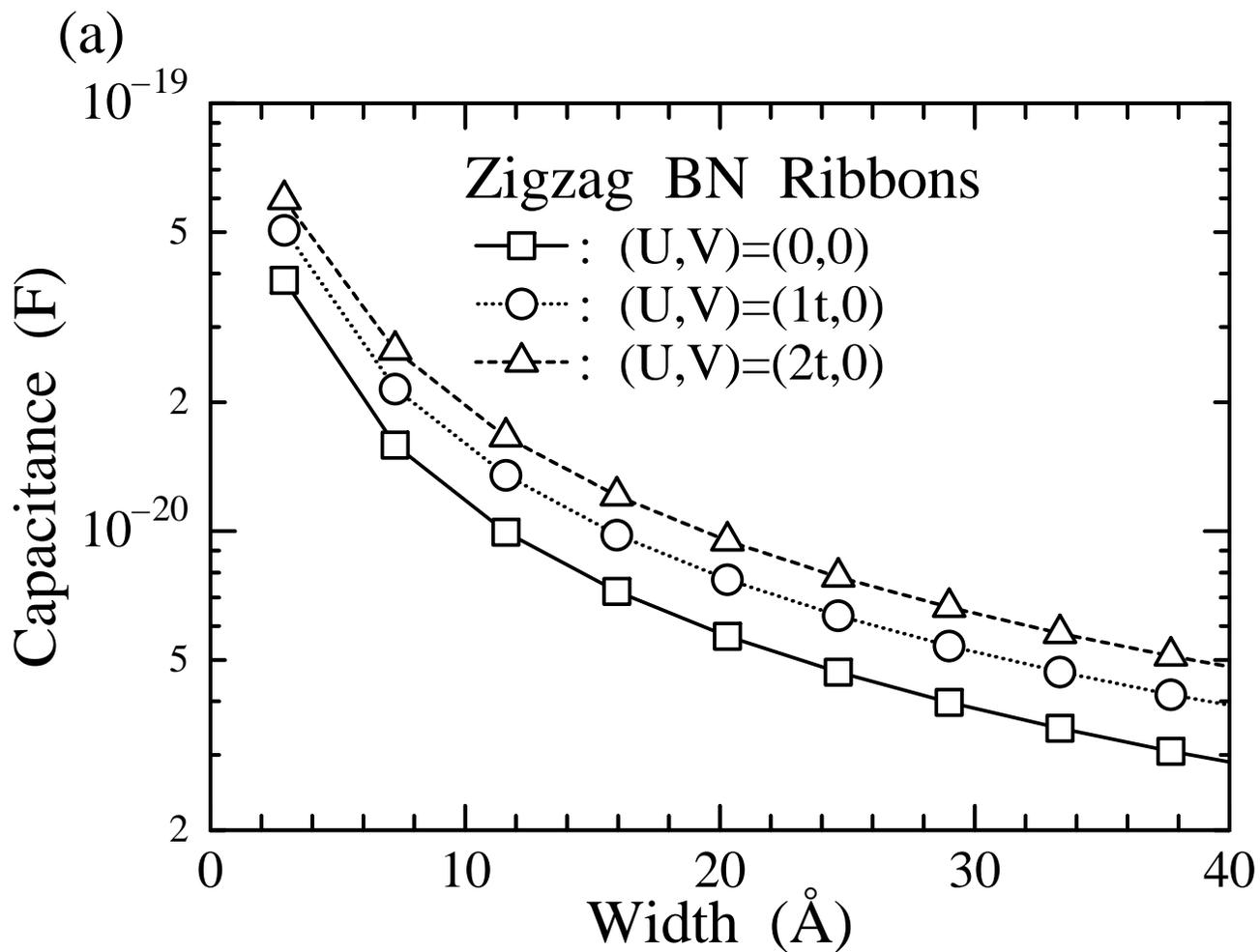
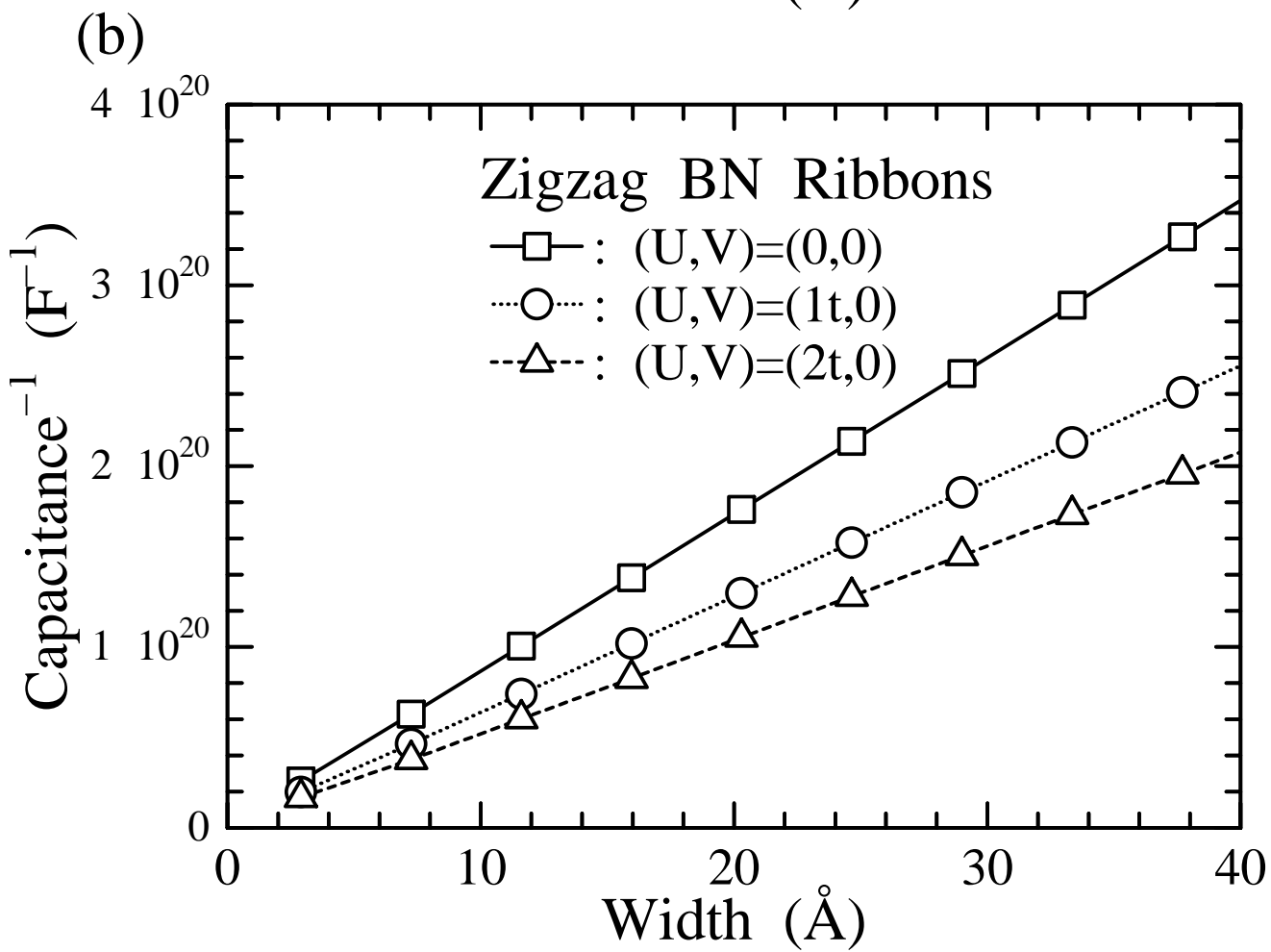
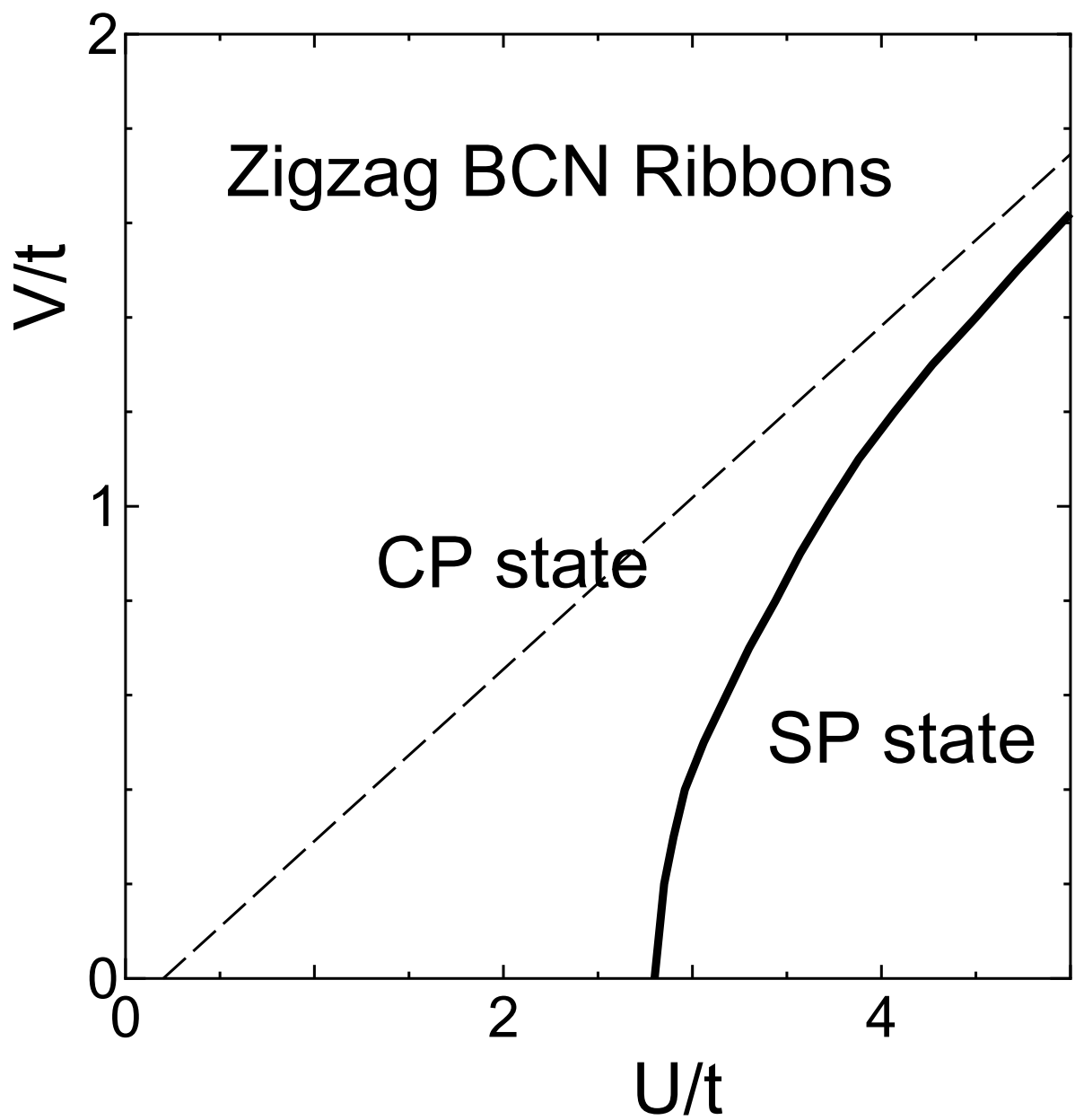
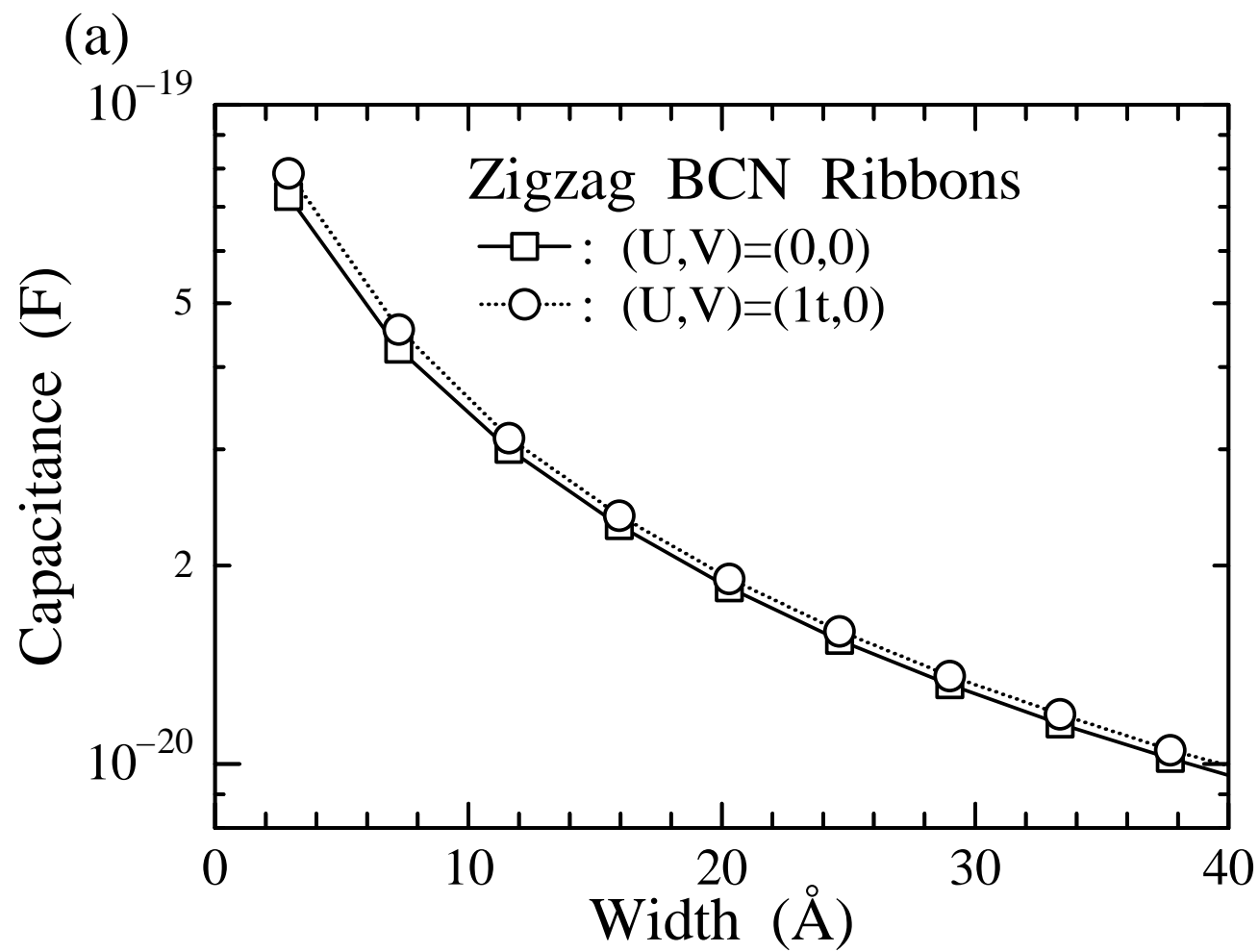
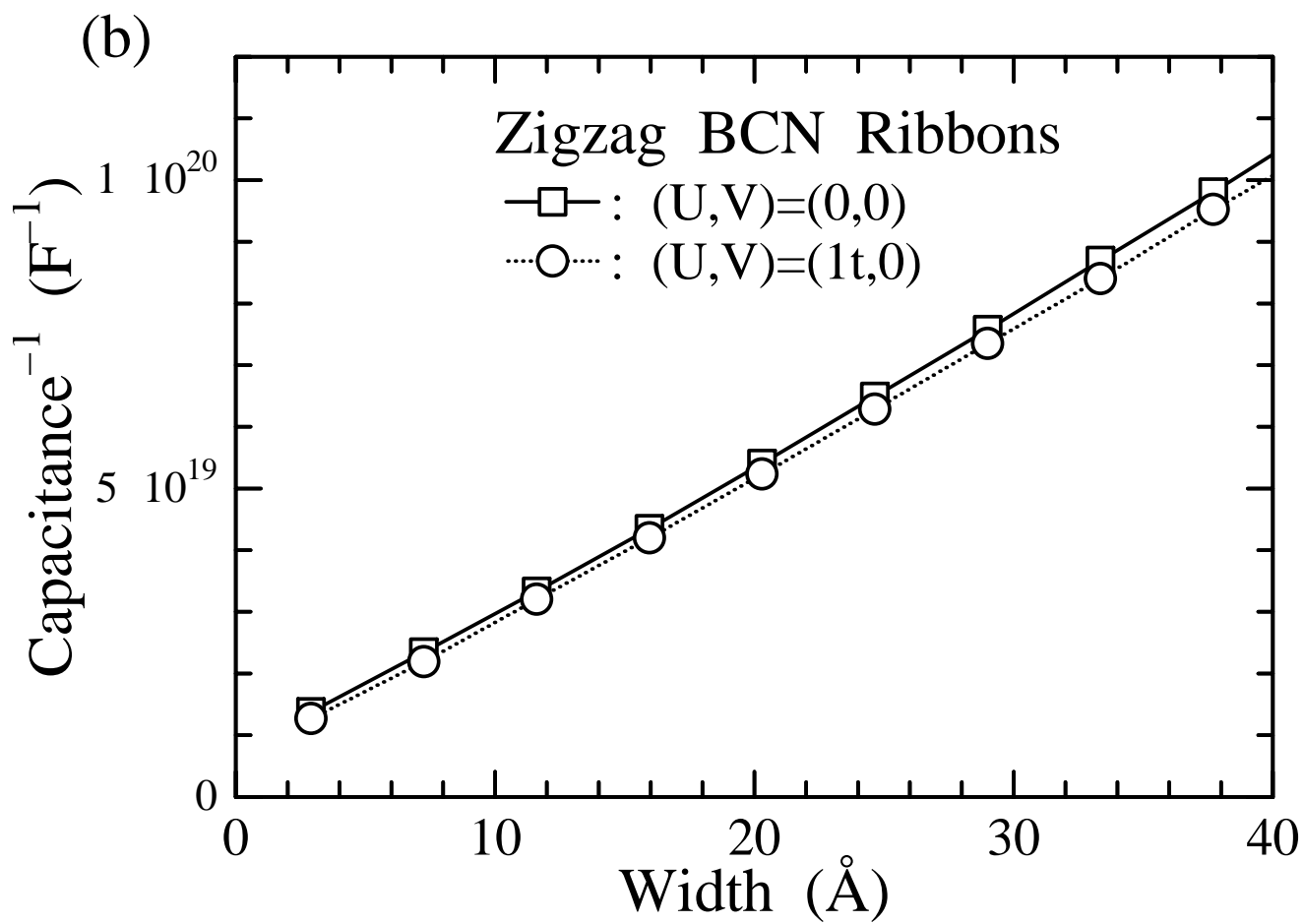
\end{document}